\documentclass{aa} 

\usepackage{graphicx}
\usepackage{txfonts}
\usepackage{natbib}
\bibpunct{(}{)}{;}{a}{}{,} 

\def\fermi{{\it Fermi\/}}

\def\lsim{\mathrel{\lower .85ex\hbox{\rlap{$\sim$}\raise
.95ex\hbox{$<$} }}}
\def\gsim{\mathrel{\lower .80ex\hbox{\rlap{$\sim$}\raise
.90ex\hbox{$>$} }}}
\newbox\grsign \setbox\grsign=\hbox{$>$}
\newdimen\grdimen \grdimen=\ht\grsign
\newbox\laxbox \newbox\gaxbox
\setbox\gaxbox=\hbox{\raise.5ex\hbox{$>$}\llap
     {\lower.5ex\hbox{$\sim$}}}\ht1=\grdimen\dp1=0pt
\setbox\laxbox=\hbox{\raise.5ex\hbox{$<$}\llap
     {\lower.5ex\hbox{$\sim$}}}\ht2=\grdimen\dp2=0pt
\def\gax{\mathrel{\copy\gaxbox}}
\def\lax{\mathrel{\copy\laxbox}}

\begin{document}

 \title{The
exotic fraction among unassociated \fermi\ sources}

   \author{N. Mirabal\inst{1}, D. Nieto\inst{1}, and S. Pardo\inst{1}
          }
\offprints{N. Mirabal, \email{mirabal@gae.ucm.es}}
\institute{Dpto. de F\'isica At\'omica, 
Molecular y Nuclear, Universidad Complutense de 
Madrid, Spain
           }

  \abstract
    {}
 {Revealing the nature of unassociated high-energy ($\gax 100$ MeV)
$\gamma$-ray sources remains a challenge
35 years after their discovery. Of the
934 $\gamma$-ray sources at high Galactic latitude ($|b| \geq 15^{\circ}$)
in the First \fermi-LAT catalogue (1FGL),
316 have no obvious associations at other wavelengths. We present an improved
method that automatically ranks counterparts based on their similarity
with identified $\gamma$-ray sources. 
}
{
In this paper, we apply the $K$-means unsupervised classification algorithm
to isolate potential counterparts for 18 unassociated \fermi\ sources 
contained within 
a $\sim 3000$ deg$^{-2}$ `overlap region' of the sky intensively covered 
in radio and optical wavelengths. 
}
 {
Combining our results with previous works, we reach
potential associations for  119 out of the 128  \fermi\ sources
within said region. 
If these associations are correct, we estimate that less than
$20$\% of all remaining unassociated 1FGL sources at 
high Galactic latitude ($|b| \geq 15^{\circ}$) might host
`exotic' counterparts distinct from
known classes of $\gamma$-ray emitters. Potentially 
even these outliers could be explained by high-redshift/dust-obscured
analogues of the associated sample or by intrinsically faint radio objects. 
}
{Although our  estimate of exotic sources leaves 
some room for novel discoveries, it severely restricts  the 
possibility of detecting  dark matter subhaloes and other unconventional types
of $\gamma$-ray emitters in the 1FGL. 
In closing, we argue  that the identification of \fermi\ sources at the 
low end of the flux density distribution 
will be a complex process that might only be achieved through a 
clever combination of
refined classification algorithms, 
multi-wavelength efforts, and dedicated optical spectroscopy. 
}


\headnote{}
\titlerunning{}
\authorrunning{Mirabal, Nieto \& Pardo}

\maketitle

%

\section{Introduction}
The nature of high-energy ($\gax 100$ MeV)
$\gamma$-ray sources lacking counterparts at other wavelengths
  remains an enigma
decades after their discovery. 
Early work on individual  unassociated $\gamma$-ray
sources
dates back to days of the  Cos B satellite \citep{julien}.
The mystery intensified with the discovery of 271  
$\gamma$-ray sources by the EGRET instrument aboard
the {\it Compton Gamma-Ray Observatory} \citep{hartman}.
To date, about
130 of these EGRET sources remain unassociated with about
half located at high Galactic latitude \citep{muk2,thompson}.
The latest source count in the 100 MeV to 100 GeV
energy range produced by the Large Area Telescope (LAT) 
instrument 
on board the \fermi\ Gamma-ray Space Telescope has expanded 
the number of persistent high-energy $\gamma$-ray sources
at high Galactic latitude ($|b| \geq 15^{\circ}$)
to 934
\citep{abdo}. But despite its superb 
angular and energy resolution relative to EGRET, only 618 of these sources 
are confidently associated. This leaves 316 unassociated
\fermi\ sources at $|b| \geq 15^{\circ}$.

While multi-wavelength strategies have evolved considerably, the
difficulties inherent  
to the identification process  
of $\gamma$-ray sources ({\it i.e.} arcminute-scale error regions) 
continue to afford some 
room for speculating about the nature of the unassociated population.
\citet{montmerle} estimated that a combination of supernova
remnants, OB groups, and even H II regions could account
for a number of the as-yet 
unassociated $\gamma$-ray sources 
\citep[see also the ground-breaking work done by][]{morrison}. 
Off the Galactic plane, 
there are statistical 
hints of $\gamma$-ray emission from stacked galaxy clusters \citep{scharf}. 
Recently, a number of
authors have pointed out that even rarer $\gamma$-ray emitters including 
subhaloes left behind during
structure formation could be detected by \fermi\ 
  \citep{pieri,kuhlen,buckley}.
Throughout the text, we shall use the term `exotic' 
to refer to any type of $\gamma$-ray source that is clearly distinct
from the associations described in \citet{abdo}. 

Typically, the procedure used 
to generate $\gamma$-ray source associations relies on the positional
coincidences between \fermi\ sources and 
catalogues of plausible $\gamma$-ray emitters 
categorised at other wavelengths \citep{reimer,abdo}. 
A lengthier approach to source association involves
the brute-force search for a counterpart
using deep radio, X-ray, and optical observations,
together with  spectroscopic classification of notable
objects within its $\gamma$-ray error region \citep{mirabal,mukherjee}. If
no plausible association can be produced through either method, then
the source would fall within our definition of
`exotic' as it must be lacking one or more of the
fundamental
attributes of the known classes of $\gamma$-ray emitters.
Regardless of the association method, a firm physical
link between a given $\gamma$-ray source and a counterpart
in another wavelength can only be established through 
contemporaneous temporal variability, similar
spatial morphology, or equivalent pulsation. However,
only a small fraction of $\gamma$-ray sources meets any of these
criteria. As a result, 
the large majority of current associations
are only probabilistic.  

Apart from the shortcomings of current
classification methods, our ability to associate $\gamma$-ray sources
rests on the quality of the catalogues used for that purpose \citep{abdo}.
Generally, most astronomical databases used for source
association have been constructed from
disjointed surveys with limited spatial coverage/flux limits. 
Naturally, the association
process reflects such inhomogeneities. For the casual reader,
this means that some areas of
the sky are much better covered than others. In other words,
the overall likelihood for finding a source association on
positional coincidence alone is not uniform 
across the \fermi\ sky. 

With few options left for improving our $\gamma$-ray focusing capabilities  
and catalogue coverage in the short term, the best hopes 
for identifying the rest of unassociated $\gamma$-ray sources may 
lie in a further
refinement and application of classification algorithms. A possible way forward
is to take advantage of proven classification tools borrowed from the
data mining and machine learning communities to improve searches
of individual \fermi\ LAT source fields.
A number of such algorithms has been applied in an
assortment of astrophysical
applications. \citet{ball} used decision trees
to provide star/galaxy classification for the entire
Sloan Digital Sky Survey (SDSS) 
data release. Applied
agglomerative hierarchichal clustering and
$K$-means clustering have also been used 
for spectral classification in X-rays 
\citep[see][and references therein]{hojnacki}.  

Here we describe an application of $K$-means clustering as a classifier of 
objects inside the error regions of unassociated \fermi\ objects and
the subsequent estimate of the `exotic' fraction among the catalogued 
\fermi\ population.
Given the extreme 
dust extinction and source crowding close to the Galactic plane,
we restrict our analysis to high Galactic latitude. Further,
and in order to maximise the likelihood of association, we 
turn to one of the most intensively studied areas of the sky  away from
the Galactic plane, namely the 
`overlap region' \citep{kimball}. This $\sim 3000$ deg$^{2}$ area
is defined by the overlap of four radio catalogues: Green Bank 6 cm Survey 
(GB6), Faint Images of the Radio Sky at Twenty Centimeters (FIRST), NRAO--VLA
Sky Survey (NVSS), and the Westerbork Northern Sky Survey
(WENSS), as well as 
photometry and spectroscopy 
collected by the SDSS. 

The structure of the paper is as follows: \S 2 summarises
the data selection, \S 3 describes the $K$-means classification algorithm. 
\S 4 details the results of the classification process. \S 5 discusses spectral
typing. Discussion and conclusions are presented in \S 6 and \S 7.

\section{Data description}
The `overlap region' is a $\sim 3000$ deg$^{2}$ strip around the
North Galactic Cap extending between
$7.6^{h} \lax R.A. \lax 17.8^{h}$, $+28.8^{\circ} \lax decl. 
\lax +63.2^{\circ}$
 where the FIRST (20 cm), NVSS (20 cm), WENSS (92 cm), GB6 (6 cm) radio
surveys, and the SDSS optical survey coincide. This region
is well outside of the Galactic plane ($|b| \gax 25^{\circ}$) and provides
an ideal location to identify high-latitude $\gamma$-ray sources. 
A full description of the limits of each survey as well as
the respective restrictions in sky coverage 
are defined in \citet{kimball}. 

\begin{figure}
\hfil
\includegraphics[width=3.1in,angle=0.]{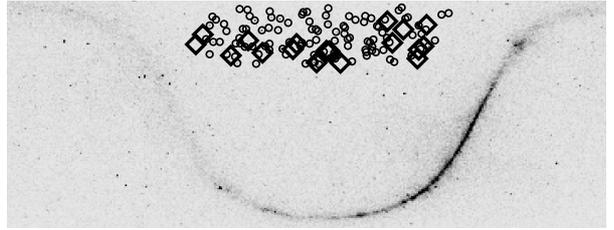}
\hfil
\caption{\fermi\ LAT all-sky map for energies $> 10$ GeV in equatorial projectio
n. 
Circles indicate associated sources. Large diamonds mark unassociated
\fermi\ sources. The
marker sizes have been
greatly exaggerated for easier visualisation. 
The footprint of the `overlap region' is outlined  
by the locations of \fermi\ sources. 
The continuous strip of $\gamma$-ray emission tracks the
Galactic plane. 
}
\label{figure1}
\end{figure}

The First \fermi-LAT
catalogue (1FGL) consists of 1451 sources characterised in the
100 MeV -- 100 GeV energy range \citep{abdo}.
Within the `overlap region', we have
identified 128 1FGL sources \citep{abdo} that are 
simultaneously covered by all four radio surveys and the SDSS
photometric survey. 
A total of 110 sources are paired with plausible associations
in the 1FGL \citep{abdo} or the First LAT AGN Catalogue  
\citep[1LAC,][]{abdo2}. The remaining  
18 correspond to sources designated as
unassociated. Figure \ref{figure1} shows the 
distribution of these sources and a general footprint of the `overlap region'. 

The sample of associated sources comprises 61 BL Lacertae objects (BL Lacs),
41 flat-spectrum radio quasars (FSRQs), and 8 objects classified as active
galactic nuclei (AGN) of rare or unknown type. 
For practical purposes, we treat these 110 associated sources
as the main source of 
training and testing sets for the $K$-means algorithm 
that will be described in greater detail 
later.

Since it is not our intention to reinvent matching algorithms for
radio surveys,
we used the procedure introduced by \citet{kimball} to ensure physically-real
matches. After collecting the associated sources, we searched for 
their respective radio counterpart in the FIRST catalogue \citep{becker}. 
We then positionally matched the FIRST location to its closest WENSS 
detection using a 
30\arcsec\ matching radius around the FIRST position. Next, we 
matched the FIRST detection to GB6 using 
a radius of
70\arcsec\ search radius. 
Once the matches were completed, the actual spectral index $\alpha$ for each
radio counterpart detected in at least two frequency bands was calculated
as $S_{\nu} \propto \nu^{\alpha}$.

One of the concerns associated with matching algorithms applied to 
radio catalogues with widely different angular resolutions is
the possibility that the sample might be contaminated by
coincidental  physically-unrelated sources. \citet{kimball}
have used the distribution of 
angular distances to the nearest neighbour source 
to find the fraction of matches from the FIRST, WENSS, and GB6 which
are physically real. For the samples used here, the estimated efficiencies
are FIRST-WENSS ($\geq 92$\%), FIRST-GB6 ($\geq 79$\%), and FIRST-SDSS 
($\geq 95$\%) respectively. 
Thus, we can safely assume that at least $\geq 79$\% of all
the radio sources are properly matched. 

\begin{figure}
\hfil
\includegraphics[width=3.1in,angle=0.]{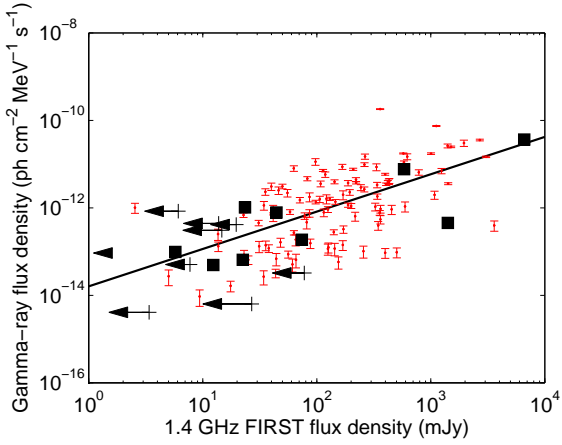}
\hfil
\caption{\fermi\ $\gamma$-ray flux density vs. 1.4 GHz FIRST flux
density ($S_{1.4}$) within the `overlap region'. 
Small dots ({\it red}) represent associated \fermi\ sources. Black squares
indicate new associations produced by the $K$-means 
classification algorithm. Arrows indicate radio upper limits for 
unassociated \fermi\ sources. The line follows an indicative fit 
$f_{\gamma} \propto S_{1.4}^{0.85}$. 
}
\label{figure2}
\end{figure}

Internally, we further validate the reported efficiency of the matching
procedure of \citet{kimball} within our sample.   
All of the 110 associated \fermi\ sources have a 1.4 GHz FIRST
counterpart brighter than 2.5 mJy, 95\% are detected by WENSS to
a limiting flux of 18 mJy,
and roughly 93\% show a GB6 source brighter than 18 mJy.
The {\it disappearance} of WENSS and GB6 counterparts
occurs predominantly for BL Lac associations at the faint end of the
FIRST radio density distribution ($S_{1.4} \lax 31$ mJy). But for
the most part, the radio regime excels at capturing the
non-thermal emission from $\gamma$-ray sources 
\citep[see also][]{kovalev,giroletti,mahony,ghir}.
Figure \ref{figure2} shows the distribution of FIRST radio fluxes
for the associated \fermi\ sources.

\section{Classification algorithm}
$K$-means is a multivariate, iterative method that  
automatically finds $K$ 
`natural' clusters in a specific dataset.  In its 
simplest form, each object 
is assigned to the cluster with the nearest cluster
centroid \citep{macqueen,hojnacki}. 
The partition process is repeated automatically until convergence has been 
reached {\it i.e.} no object reassignments are performed to a different 
cluster. 
Since our aim is to locate potential associations 
within the sample of unassociated \fermi\ sources, we considered
three input variables: the radio 
spectral index $\alpha_{92}$ between WENSS (326 MHz)
and FIRST (1.4 GHz), the radio spectral index $\alpha_{6}$ between 
FIRST (1.4 GHz) and GB6 (4.85 GHz), and the $\gamma$-ray photon 
spectral index $\Gamma$ derived 
in the 100 MeV-100 GeV range \citep{abdo}. 

\begin{figure}
\hfil
\includegraphics[width=3.1in,angle=0.]{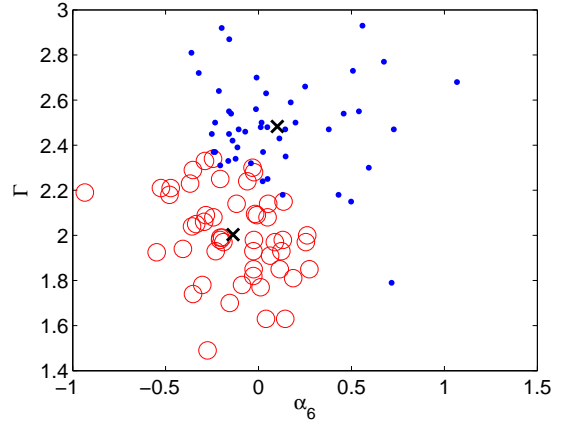}
\hfil
\caption{\fermi\ photon index $\Gamma$ vs. spectral index
$\alpha_{6}$ between 1.4 and 4.85 GHz for associated \fermi\ sources.
Blue ({\it filled circles}) and red ({\it open circles})
markers represent two distinct clusters automatically identified by $K$-means.
The X symbols mark the centroid  of each cluster.
}
\label{figure3}
\end{figure}

\begin{figure}
\hfil
\includegraphics[width=3.1in,angle=0.]{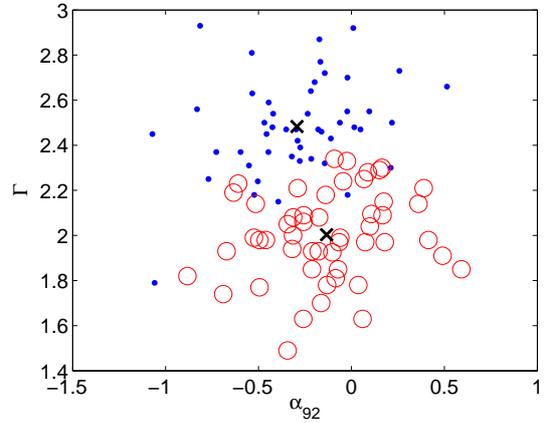}
\hfil
\caption{\fermi\ photon index $\Gamma$ vs. spectral index
$\alpha_{92}$ between 326 MHz and 1.4 GHz for associated \fermi\ sources. 
Blue ({\it filled circles})
and red ({\it open circles}) markers
represent two clusters automatically identified by $K$-means.
The X symbols mark the centroid of each cluster.
}
\label{figure4}
\end{figure}

We know {\it a priori} that there are two `natural' clusters in the
associated sample: BL Lacs with an average photon
index of $\Gamma = 2.18 \pm 0.02$ and FSRQs with an average
$\Gamma = 2.48 \pm 0.02$ \citep{abdo5}. 
Thus, the $K$-means algorithm was initially performed on
the associated \fermi\ sources assuming
$K=2$ as an input parameter.  
Figures \ref{figure3} and \ref{figure4} 
show the final separation into two distinct clusters. 
The unsupervised algorithm does a superb job  
separating spectroscopically distinct FSRQs ($\Gamma \sim 2.48$) from 
BL Lacs ($\Gamma \sim 2.01$). However, there is a  clear 
conflict region at
the boundary of the clusters that results in
$\sim$ 30\% {\it false positives} for spectroscopic BL Lacs 
and $\sim$ 33\% {\it false positives} for spectroscopic 
FSRQs. While the classification
is not perfect, the results  demonstrate the 
effectiveness of unsupervised classification algorithms in this context.

To aid in the identification of possible counterparts within
the error regions of 
unassociated \fermi\ sources, 
we need to define a locus 
that can help us
recognise the position of possible associations in $\Gamma$--$\alpha$
space. We accomplish this by splitting the
associated sample into 100 random training (70\% of the total) and testing 
(30\% of the total) sets. For each individual testing set, we performed
the $K$-means algorithm to automatically find the cluster centroids of that 
particular set. Subsequently, its 
associated testing set allowed us to quantify the performance of
the algorithm  by counting the number of {\it false positives} as a function
of threshold around the centroid.  Additional refinement was achieved with 
template radio spectral indices culled from
possible contaminants expected in \fermi\ source fields including 
quasi-stellar objects (QSOs) and radio galaxies \citep{mirabal}. 

Generally, we found that the $K$-means algorithm is a powerful \fermi\ 
classifier using a locus with a narrow dispersion ($\sim 1.5 \sigma$) around 
the centroids of the distribution. Larger loci  
start to capture outliers that might include radio AGN outbursts
with steep radio spectral indices $|\alpha| \gax 0.5$. 
At larger distances from the centroids, we also start to see a 
degeneracy with
radio galaxies and certain types of non-blazar AGNs 
in the $\Gamma$--$\alpha$ space. By picking a restrictive 
locus with a small dispersion, $K$-means only classifies the core of
the distribution ($\sim$ 75\% of 
the objects) while missing or filtering the outliers of
a particular data set. However, in return, the algorithm optimises the 
classification process by reducing the 
number of {\it false positives} ($\lax 20$\%).

\section{Results}
Once the best locus parameters were determined, we applied the $K$-means 
algorithm to objects found inside 
the error regions of each of the 18 unassociated \fermi\ sources.  
In order to gather a full census of possible candidates, we 
consider all the radio objects within the 99.7\% confidence 
location contour for each source. 
The 99.7\% confidence level radius $r_{99.7}$ is related to 
the 95\% confidence level radius $r_{95}$ derived by \citet{abdo}
through
$r_{99.7} = 1.39 \times r_{95}$. 

Taking radio
emission as a proxy of $\gamma$-ray emission, 
we start with the positions of 
all FIRST radio detections within each $r_{99.7}$ error region. We 
then proceed to
match each individual FIRST location with the GB6 and WENSS catalogues
as detailed in \S 2. 
Finally, we derive $\alpha_{6}$ and $\alpha_{92}$ radio spectral indices 
for any object 
detected in at least two radio frequency bands. The process  
is very fluid as radio sources
are typically sparse. On average there are 90 sources per deg$^{2}$
down to the FIRST flux limit and even fewer in the GB6 and 
WENSS.

Next, each object with a measured spectral 
index is compared to the locus found using
the associated sample. This is a two-step procedure
that initially ranks the radio objects within the \fermi\ error region
according to their Euclidean
distance to the nearest cluster 
centroid (objects with the smallest distance are ranked at the top). 
It then  flags each object either as
 a possible association (if a radio candidate
lies within the locus) or
unassociated (outside the locus). 
In total, the automated $K$-means algorithm 
returned possible counterparts for 8 out of 18 unassociated \fermi\ sources
in our sample. Table \ref{table1} lists
the 8 sources with their respective AGN association. 
In the case of 1FGL J0942.1+4313, we were not
able to perform the $K$-means algorithm as none
of the FIRST sources within its 99.7\% error contour 
had an equivalent WENSS or GB6 counterpart. 
Unassociated sources 
without any apparent counterparts are summarised 
in Table \ref{table2}
(see appendix for further details).

\section{Spectral typing}
Optical spectroscopy is arguably 
the most powerful tool to 
accomplish the classification of
objects within a particular $\gamma$-ray 
error contour \citep{mirabal,mukherjee}.
However, manually sorting through dozens of spectra is
painstaking and difficult. The $K$-means algorithm eases the 
identification process by ranking all the objects within
the error contour. 

We take advantage of the $K$-means results
to search for publicly available optical
spectra at the position of each new association
 listed in Table \ref{table1}.
In addition to deep optical photometry, the SDSS is carrying out
an impressive SDSS Spectroscopic Survey that will eventually 
obtain calibrated spectra for about one million objects
with a spectral wavelength range 3800--9200 \AA\ and
a resolution of 1800 \citep{stoughton}. 
 After sifting through this massive sample, 
we find that 4 out the 8 associated sources
have corresponding SDSS spectra. Two (CRATES J101811+354229 and
3C 345) are spectroscopically 
classified as FSRQs at $z=1.228$ and $z=0.588$ respectively.
Two additional sources (FIRST J113812.1+411353 and 1RXS J125716.0+364713)  
look like good BL Lac candidates without obvious redshifts. 

For completeness, we also combed the SDSS Spectroscopic survey for
spectra at the positions of all FIRST sources contained within the
99.7\% error regions for the remaining 10
unassociated \fermi\
sources. Most of the matching spectra are either run-of-the-mill 
QSOs or galaxies. However, we have localised an additional 
BL Lac associated with FIRST J124946.7+370748 inside
the error contour of 1FGL J1249.8+3706. 
Accordingly, we add the latter to the associated column.
We note that 
FIRST J124946.7+370748 failed to be associated by the $K$-means algorithm as
it is only detected in a single frequency 
with a 1.4 GHz FIRST flux density of 5.75 mJy.   
SDSS optical spectra of the associated objects 
are shown in Fig. \ref{figure5}. Additional
notes on individual objects are given in the appendix.

\section{Discussion}
We have successfully applied the $K$-means classification algorithm 
to 18 unassociated \fermi\ sources within the `overlap region'. 
The algorithm trained on associated sources enables the potential 
classification
of 8 new \fermi\ sources. 
Adding these to one additional  
source spectroscopically associated with 1FGL J1249.8+3706  
reduces the number of 1FGL
unassociated sources in the `overlap region' from 18 to 9. 
Proper accounting indicates that 119 out of 128 \fermi\ sources (93\% of all
\fermi\ sources) are 
associated within said region. In contrast,  
outside the `overlap region', 
only 508 out of 806 ($\sim$ 63\%) 
\fermi\ sources  at $|b|\geq 15^\circ$
have been associated. Assuming that the 1FGL sky coverage 
is nearly uniform outside the
Galactic plane ($|b|\gax 10^\circ$), where diffuse $\gamma$-ray 
emission is less prominent, 
we find that the percentage of unassociated 
\fermi\ sources in the `overlap region' suggests that $\lax$ 20\%
of all remaining unassociated 1FGL sources at $|b|\geq 15^\circ$ 
might host new types of $\gamma$-ray 
sources. As noted earlier, such discrepancy 
is partially due to better coverage and
more complete catalogues in the northern sample. 

However, we are faced with a clear puzzle: 
What kind of counterparts are hiding among 
the 9 `exotic' outliers?.
To explore this question we plot in Figure \ref{figure2} 
the \fermi\ $\gamma$-ray flux density from
\citep{abdo} versus the 1.4 GHz FIRST flux density for 
the 9 unassociated \fermi\ sources. In the same plot, we
include the complete sample of associated \fermi\ sources within
the `overlap region'. For the unassociated sources, we assign a radio 
upper limit from the flux density of 
the brightest FIRST radio source
within the error region that is not spectroscopically classified
either as a galaxy or QSO (see Table \ref{table2}).

While there is significant scatter   
between the two quantities, we notice that in general
brighter $\gamma$-ray sources tend to be brighter in radio. A simple
fit $f_{\gamma} \propto S_{1.4}^{0.85}$ has been drawn through the
points to show an indicative general trend. Interestingly, our best-fit slope
at 1.4 GHz is identical to the result of \citet{ghir} 
based on an entirely different set of associated 
\fermi\ sources acquired at 20 GHz. 
With the notable exception of 1FGL J1527.6+4152 at 77 mJy, 
the \fermi\ gamma-ray flux density and radio upper limits   
of the 9 unassociated sources 
tend to lie at the faint end of the distributions
($S_{1.4} \lax 27$ mJy). The observed scatter of
radio flux values admits that
the actual associations for these outliers 
could be high-redshift or dust-obscured analogues of sources already
present in the 1FGL. 

The latter reasoning is weakly reinforced by the $K$-means
association of 1FGL J0753.1+4649 (without an apparent 
optical counterpart) and the fact
that an important fraction of the remaining 
FIRST radio sources detected within the 
error regions of the 9 unassociated \fermi\ sources lack an SDSS optical 
counterpart to a limit of $r > 23.1$. Alternatively, the  
outliers could have intrinsically 
fainter radio counterparts. Either requirement could be met with the
known population of $\gamma$-ray emitters in the 1FGL without
invoking new types of sources. For instance, although rare, it is possible
that at least one high-latitude
radio-quiet pulsar could be hiding 
within the `overlap region' \citep{mirabal2,halpern1,abdo6}.

\section{Conclusions}
It is important to emphasise once again that 
the associations reported here
are not final identifications but rather statistically significant matches. 
While the matching procedure for radio sources is not perfect, 
the conclusions appear firm since there is
little contamination from
physically-unrelated sources \citep{kimball}.
Ultimately, the goal of 
this paper is to introduce an algorithm that can facilitate the search for 
counterparts within {\it Fermi} error circles by ranking their similarity with 
previously well-identified Fermi counterparts. Empirically, this method 
is already implemented in manual searches but the algorithm presented
here automates 
the process. The multifrequency observer can now start to study
a certain $\gamma$-ray error circle
with a possible order of priorities.
Strictly speaking the detection of
contemporaneous variability, pulsations, or spatial extent 
are the only paths to directly prove  
a physical connection. Unfortunately,
except for radio-loud pulsars and a fraction of the brightest \fermi\ 
AGN, a firm identification
might have to wait for the capabilities of future
very high-energy ($> 100$ GeV) 
$\gamma$-ray arrays that could start to probe correlated variability with
more sensitive instruments \citep{hinton}.

Clearly, the association of \fermi\ sources at the lower end  
of the flux distribution looms
as a pressing issue not only to complete the census of 
unassociated sources but also to help pinpoint objects 
responsible for the unresolved, extragalactic diffuse 
$\gamma$-ray emission at energies above 200 MeV \citep{abdo3}.
While the absence of obvious counterparts lets us
entertain the possibility that 
a handful of unconventional 
objects such as dark matter subhaloes are powering certain 
\fermi\ sources, the approach presented here represents a
first attempt to cap
their actual fraction. 
Yet, we caution that the pay-off for finding a 
dark matter signal among unassociated \fermi\ sources 
would be so critical for the progress of 
fundamental physics that it merits continued efforts 
to identify every single  source 
in the 1FGL. 

Possibly, the ideal association process for the outliers 
would involve collecting spectroscopy
for the totality of sources within their error regions.  Presumably, 
such coverage
will be hard to accomplish with current instrumentation, especially 
given the difficulties at the
low end of the flux distribution across bands. 
As shown here, classification algorithms could play a crucial 
role in isolating  clusters of truly 
`exotic' objects. However, the variables and locus definition
presented here should not be taken as universal models.   
In the future, further refinement
of the algorithm will take place 
when additional coverage in various wavelengths is 
completed and a final classification of all  the sources is reached.
Once final classifications are achieved, additional 
work shall explore alternative 
classification algorithms as well as an additional input 
variables. One foreseeable limitation for the $K$-means
algorithm appears to be 
the flux depth of existing surveys, but that could be improved
with dedicated observations.  
We close by noting that the estimated fraction 
could be further restricted with dedicated multi-wavelength 
efforts and pulsation searches in the `overlap region'.

\begin{center}
\begin{figure*}
\hfil
\includegraphics[width=1.4in,angle=0.]{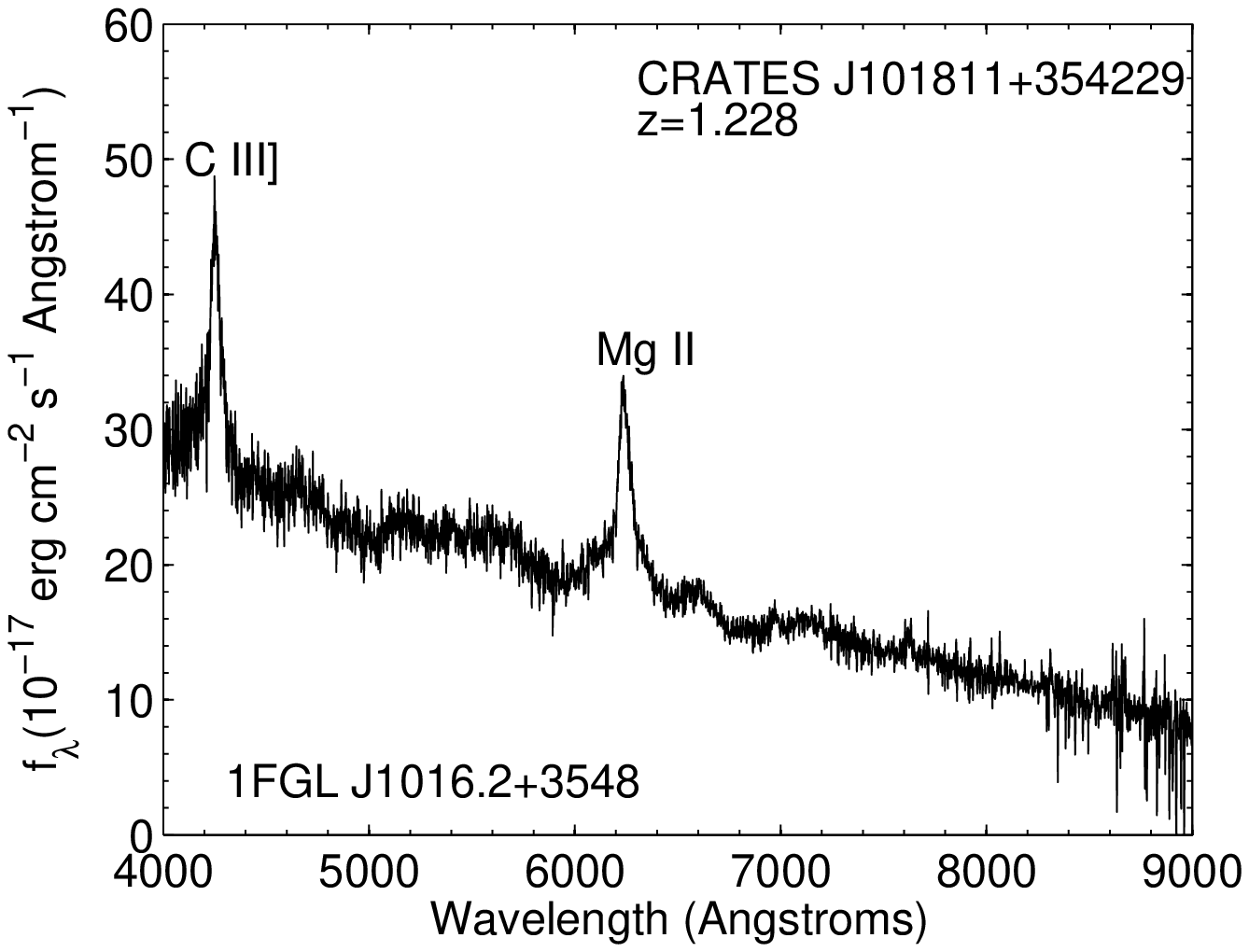}
\includegraphics[width=1.4in,angle=0.]{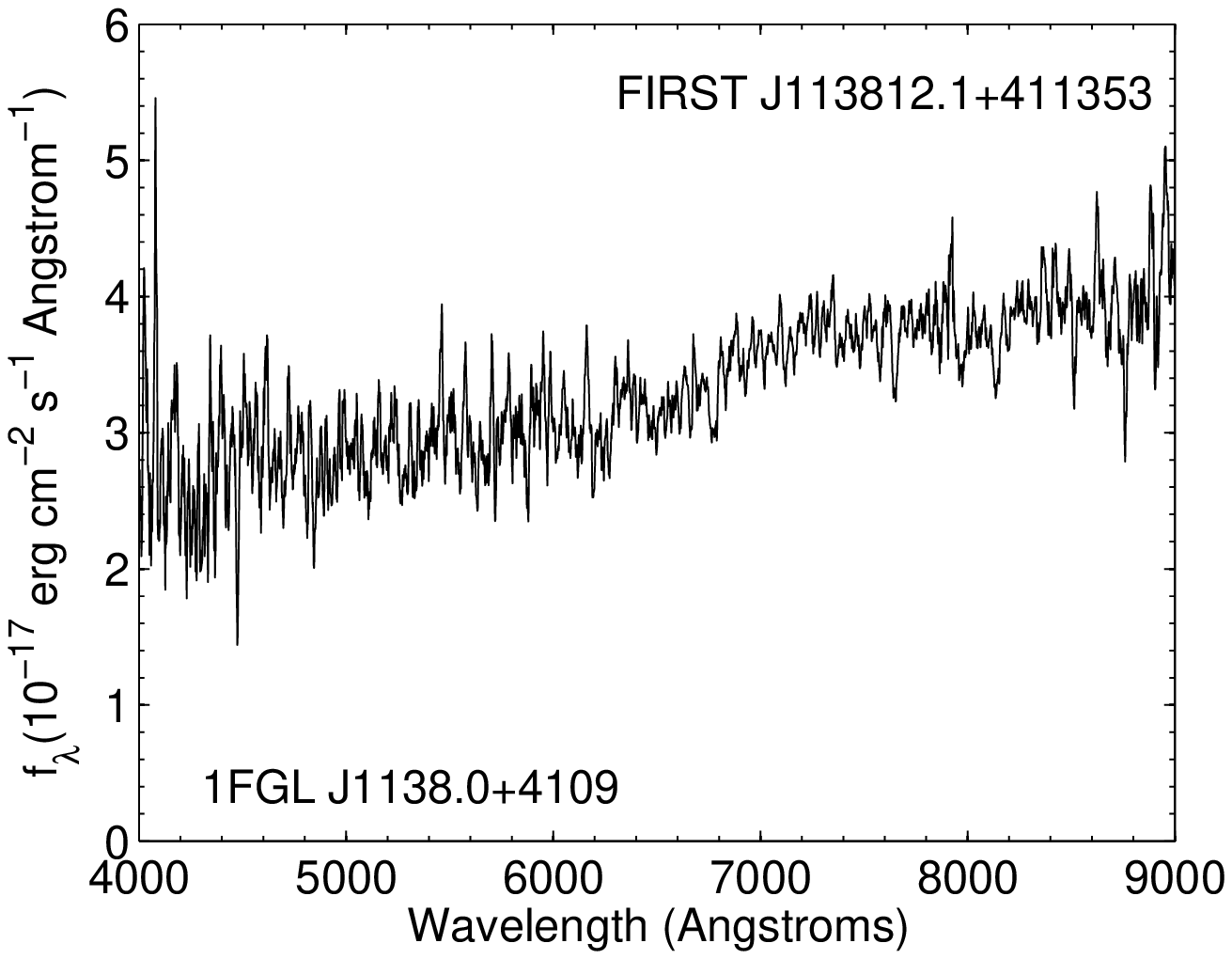}
\includegraphics[width=1.4in,angle=0.]{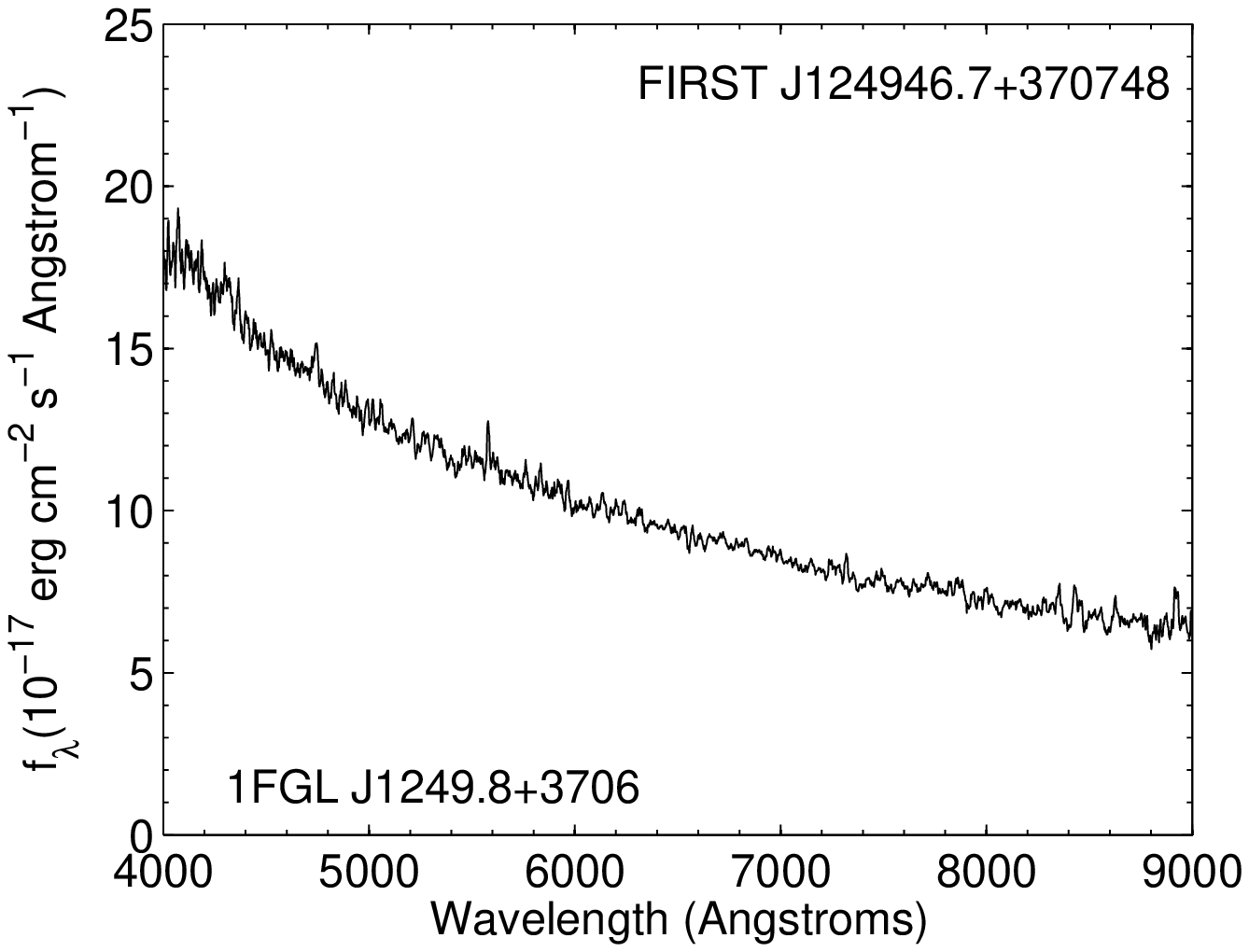}
\includegraphics[width=1.4in,angle=0.]{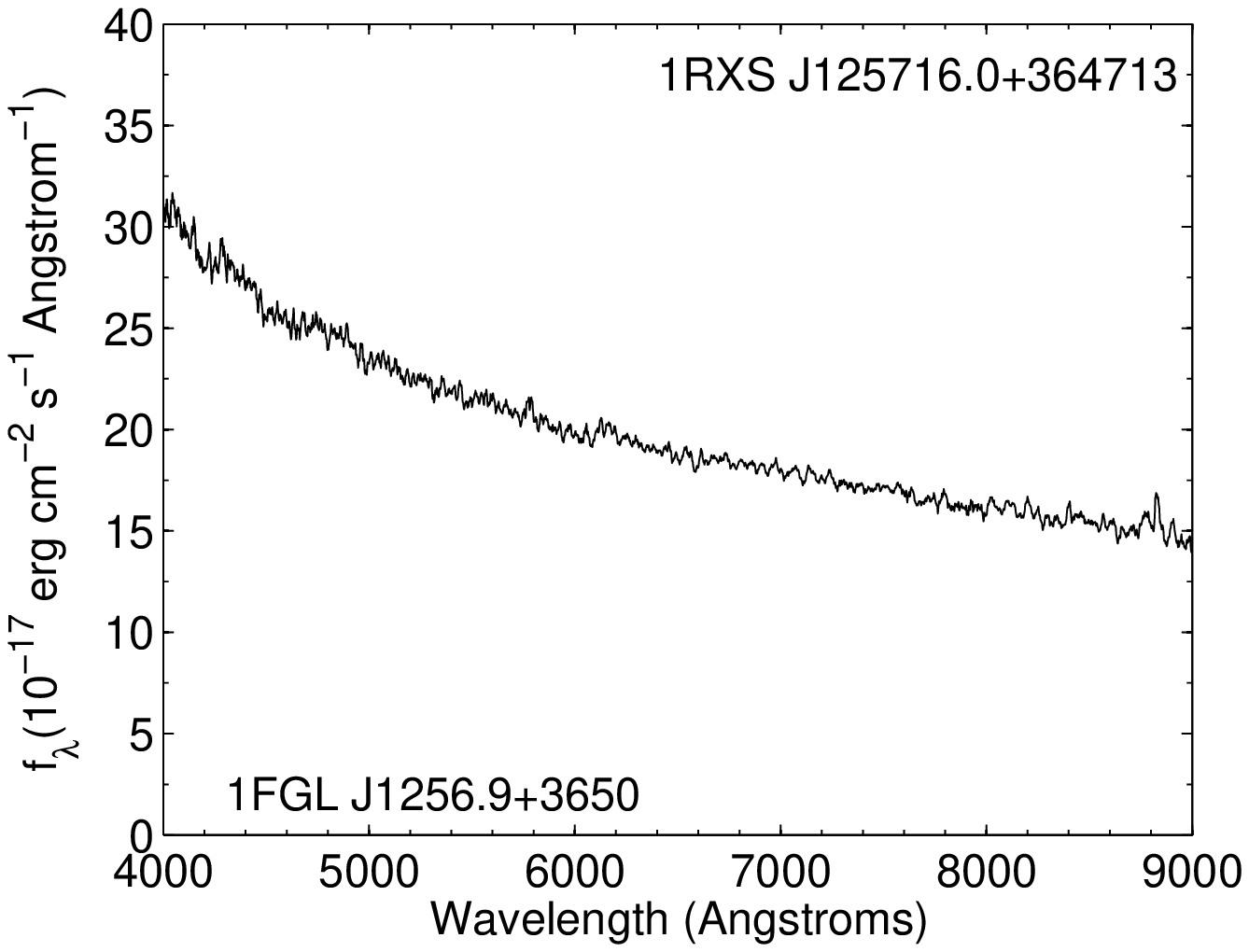}
\includegraphics[width=1.4in,angle=0.]{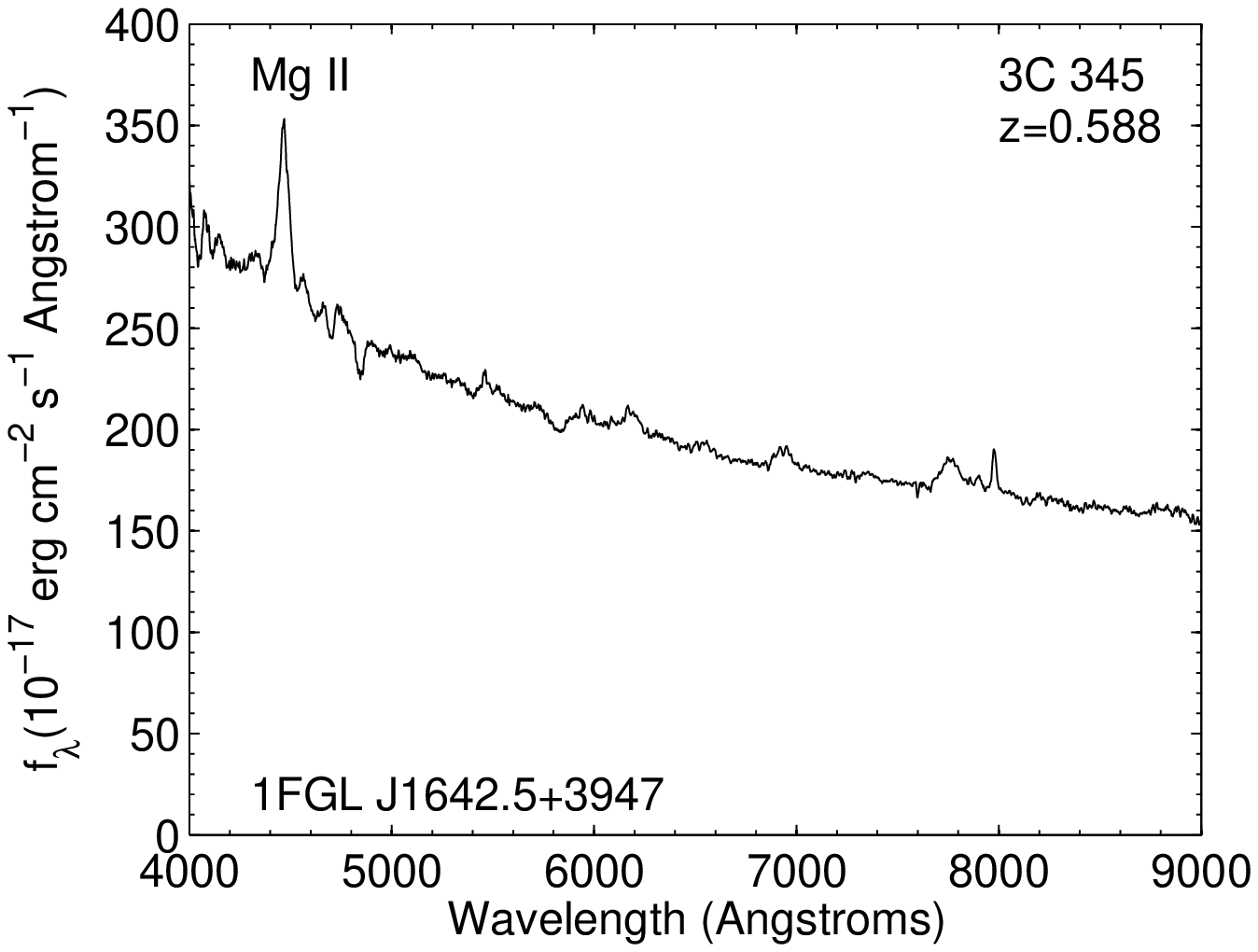}
\hfil
\caption{
SDSS spectra of newly associated sources. 
}
\label{figure5}
\end{figure*}
\end{center}

\begin{acknowledgements}
N.M. gratefully acknowledges support from the Spanish Ministry of Science
and Innovation through a Ram\'on y Cajal fellowship.
We thank Jules Halpern for sharing some of his spectroscopic wizardry with us. 
We acknowledge useful correspondence with 
Phil Gregory, David J. Thompson and an anonymous reader. 
We also 
acknowledge support from the 
Consolider-Ingenio 2010 Programme under grant MULTIDARK 
CSD2009-00064. This paper made use of data from the SDSS.
Funding for the SDSS and SDSS-II has been provided by 
the Alfred P. Sloan Foundation, the Participating Institutions, the 
National Science Foundation, the U.S. Department of Energy, the National Aeronau
tics and Space Administration, the Japanese Monbukagakusho, the Max Planck Socie
ty, and the Higher Education Funding Council for England. The SDSS Web Site is h
ttp://www.sdss.org/.

   The SDSS is managed by the Astrophysical Research Consortium for the Partici
pating Institutions. The Participating Institutions are the American Museum of N
atural History, Astrophysical Institute Potsdam, University of Basel, University
 of Cambridge, Case Western Reserve University, University of Chicago, Drexel Un
iversity, Fermilab, the Institute for Advanced Study, the Japan Participation Gr
oup, Johns Hopkins University, the Joint Institute for Nuclear Astrophysics, the
 Kavli Institute for Particle Astrophysics and Cosmology, the Korean Scientist G
roup, the Chinese Academy of Sciences (LAMOST), Los Alamos National Laboratory, 
the Max-Planck-Institute for Astronomy (MPIA), the Max-Planck-Institute for Astr
ophysics (MPA), New Mexico State University, Ohio State University, University o
f Pittsburgh, University of Portsmouth, Princeton University, the United States 
Naval Observatory, and the University of Washington.

\end{acknowledgements}

{}

\bibliographystyle{aa}

\begin{table*}
\begin{center}
\caption{\fermi\ sources with associations} 
\label{table1}
 \begin{tabular}{ccccccc}
 \hline
1FGL Name & Assoc. AGN & $S_{1.4}$ (mJy) & Spectral class &
 $z$ & $K$-means ass.?\footnotemark[1] & Spectr.? \footnotemark[2]\\
\hline
1FGL J0753.1+4649 & FIRST J075339.9+464824   & 12.36 & --
& -- & Y  & N\\
 & FIRST J075309.7+465155    & 12.83   & --  &
-- & N  & N\\
1FGL J1016.2+3548 & CRATES J101811+354229   & 582.58 & FSRQ
& 1.228 & Y  & Y\\
1FGL J1129.3+3757 & FIRST J112903.2+375656    & 23.43   & --  &
-- & Y  & N\\
1FGL J1138.0+4109 & FIRST J113812.1+411353     & 22.44   & BL Lac?  &
-- & Y  & Y\\
1FGL J1249.8+3706 & FIRST J124946.7+370748    & 5.75   & BL Lac? &
-- & N  & Y\\
1FGL J1256.9+3650 & 1RXS J125716.0+364713    & 73.86   & BL Lac  &
-- & Y  & Y\\
 & 4C +36.22    & 716.97   & Seyfert 1  &
0.709 & N  & Y\\
1FGL J1323.1+2942  & 4C +29.48   & 1412.49\footnotemark[3]   & -- &
-- & Y  & N \\
1FGL J1642.5+3947  & 3C 345    & 6598.19   & FSRQ & 
0.588 & Y & Y\\
1FGL J1649.6+5241 & FIRST J164924.9+523515   & 44.30  & -- &  
-- & Y  & N\\
\hline
\end{tabular}\\
\begin{flushleft}
$^{1}$ ``Y'' indicates that the object has been associated by the
$K$-means algorithm.\\
$^{2}$ ``Y'' indicates that the object has been associated 
spectroscopically.\\
$^{3}$ This value represents the sum of the individual components of
a multi-component radio source.
\end{flushleft}
\end{center}
\end{table*}

\begin{table*}
\begin{center}
\caption{\fermi\ sources without associations} 
\label{table2}
 \begin{tabular}{ccc}
 \hline
1FGL Name & Radio upper limit S$_{1.4}$ (mJy) & E$_{max}$ 
(GeV)\footnotemark[1]\\
\hline
1FGL J0736.4+4053 & $\leq$ 0.82 & 42.5\\   
1FGL J0900.5+3410 & $\leq 19.64$ & 29.6\\
1FGL J0942.1+4313 & $\leq  3.38$ & 35.8\\
1FGL J1226.0+2954 & $\leq 13.78$ & 10.0\\
1FGL J1515.5+5448 & $\leq 26.89$ & 14.2\\
1FGL J1527.6+4152 & $\leq 77.77$ & 21.5\\
1FGL J1553.9+4952 & $\leq 7.72$ & 265.4\\
1FGL J1627.6+3218 & $\leq 14.68$ & 7.6\\
1FGL J1630.5+3735 & $\leq 6.09$ & 6.3\\
\hline
\end{tabular}\\
\begin{flushleft}
$^{1}$ Maximum observed energy of  \fermi\ LAT photons within
99.7\% \fermi\ error region as of 2010 May 31 UT.\\
\end{flushleft}
\end{center}
\end{table*}

\appendix

\section{Notes on individual sources}

\subsection{Associated}
{\bf 1FGL J0753.1+4649.--} Two FIRST sources are catalogued
within the 99.7\% error region. FIRST J075339.9+464824
($S_{1.4}$ = 12.36 mJy) is picked automatically by $K$-means.
Oddly it lacks any apparent optical
SDSS counterpart to a limit of 
$r > 23.1$. A possible `dark horse' candidate
is the other radio 
source FIRST J075309.7+465155 ($S_{1.4} = 12.83$ mJy) with
a measured SDSS optical counterpart
at $r = 19.86$. The latter was not detected by
WENSS or GB6 and hence it could not be evaluated by the $K$-means 
algorithm. Spectroscopy is needed to settle the issue.

\noindent
{\bf 1FGL J1016.2+3548.--} The $K$-means algorithm
selects CRATES J101811+354229 as an association.
The SDSS spectrum confirms this object as an FSRQ
at $z = 1.228$ 
(see Fig. \ref{figure5}). This source  
was first suggested as an affiliation by \citet{abdo2}.  
\citet{gregory} designate this source as a likely radio 
variable at 4.85 GHz. FIRST J101505.6+360452 is most likely 
a QSO at $z = 0.843$.

\noindent
{\bf 1FGL J1129.3+3757.--}  
FIRST J112903.2+375656 is selected as an association by $K$-means. 
No optical spectrum is available for this object.
However, the derived SDSS optical colours match the  
colour-colour placement of high-confidence 
BL Lac candidates identified by \citet{plotkin}. 
At the edge of the error region lies the intriguing 
B3 1127+380, a double-symmetric
radio source with a steep radio spectral index. 1RXS J112909.8+380847,
the brightest X-ray source
within the error region, appears to
be a coronal-emitting star.

\noindent
{\bf 1FGL J1138.0+4109.--} According to $K$-means, 
FIRST J113812.1+411353 is the most likely association. Its SDSS optical
spectrum (Fig. \ref{figure5}) 
lacks any spectral features, so we regard this object
as a possible BL Lac. This source was suggested as a tentative
BL Lac by  \citet{plotkin}.
A bright X-ray source at the edge of the error region 1RXS J113857.0+410840
 corresponds to the F star HD 101207.

\noindent
{\bf 1FGL J1249.8+3706.--} FIRST J124946.7+370748 was selected 
spectroscopically 
as a possible BL Lac counterpart for the $\gamma$-ray source. The optical
SDSS spectrum shows no obvious emission or absorption 
lines (see Fig. \ref{figure5}). 
In radio, it is only detected
by FIRST with a flux density of 5.75 mJy. It was not 
included in the object list evaluated by 
the $K$-means algorithm since it lacks WENSS and GB6 counterparts.
This object could be representative of possible associations
at the faint end of the 1FGL flux density 
distribution. 

\noindent
{\bf 1FGL J1256.9+3650.--} 1RXS J125716.0+364713 was isolated
by the $K$-means algorithm. The SDSS spectrum confirms the
absence of spectral features, which is consistent with 
a possible BL Lac object (Fig. \ref{figure5}). 
This is another source 
listed as a tentative affiliation in 
\citet{abdo2}. Another intriguing object within the
error region is 4C +36.22 a Seyfert 1 at $z = 0.709$ \citep{eracleous}.
Seyfert 1 galaxies have been recently shown as potential
$\gamma$-ray emitters \citet{abdo4}. Close follow-up observations 
of both objects 
will be needed to determine the actual counterpart. 

\noindent
{\bf 1FGL J1323.1+2942.--} 
4C +29.48 is the association selected by the $K$-means algorithm. The latter
corresponds to a multi-component radio object. There is very little
information about its corresponding optical counterpart. This same source
was first suggested as a possible affiliation by \citet{abdo2}.

\noindent
{\bf 1FGL J1642.5+3947.--} $K$-means reveals 3C 345 as the
likely association. The SDSS spectrum of this object 
confirms  it as an FSRQ at 
$z= 0.588$ (Fig. \ref{figure5}). 
The actual radio counterpart flashes the 
brightest FIRST flux density among
all 1FGL associated sources  in the `overlap region'.  This object
was also noted as a possible affiliation by \citet{abdo2}.

\noindent
{\bf 1FGL J1649.6+5241.--} FIRST J164924.9+523515 was the pick by the
$K$-means algorithm. No SDSS 
optical spectrum is available for the corresponding
optical counterpart. 
The source was recognised 
as a likely variable by \citet{gregory}. It was originally
tagged as a tentative affiliation by \citet{abdo2}.
Optical spectroscopy is needed for conclusive typing.

\subsection{Unassociated}

\noindent
{\bf 1FGL J0736.4+4053.--} The brightest FIRST radio source
within the error region is FIRST J073655.0+405351 ($S_{1.4}$ = 2.34 mJy).
The latter is listed as a radio galaxy at $z = 0.352$ according to
SDSS measurements. The only other radio source within the 99.7\% error
region is FIRST J073614.4+405326
($S_{1.4}$ = 0.82 mJy) that
reveals a blank optical field to the SDSS limit of $r > 23.1$. 
FIRST J073610.4+405940 ($S_{1.4}$ = 18.85 mJy) lies
just outside the 99.7\% error region and shows a steep 
spectral index $\alpha_{92} = -0.94$ between WENSS (326 MHz)
and FIRST (1.4 GHz).

\noindent
{\bf 1FGL J0900.5+3410.--} The brightest radio source 
within the 99.7\% error region 
FIRST J085917.2+340908 ($S_{1.4} = 55.51$ mJy) 
has been identified as a galaxy at $z = 0.55$
by \citet{hook}. The second brightest radio emitter 
FIRST J090051.4+342440
($S_{1.4} = 19.64$ mJy)
is a rather steep $\alpha_{92} = -1.24$ multi-component source. 
Other multi-component radio sources lie 
in the vicinity of the error region. 
The only contained ROSAT X-ray source  1RXS J090003.5+340905
corresponds to a radio-quiet QSO ($z = 1.937$) typed by the SDSS pipeline.
Another X-ray source catalogued by XMM-Newton XMMSL1 J090046.0+335422
has been classified as a radio-quiet 
QSO at $z=0.228$ from the SDSS spectroscopic pipeline. 

\noindent
{\bf 1FGL J0942.1+4313.--} This is the faintest
unassociated \fermi\ source in the `overlap region'. 
Only two FIRST sources lie 
within its error region. One is a galaxy: FIRST J094136.4+431752 ($z = 0.15$). 
The other
radio source FIRST J094230.5+430920   
is detected with a flux density of 3.38 mJy at 1.4 GHz 
but lacks an optical SDSS counterpart to a limit of $r > 23.1$.

\noindent
{\bf 1FGL J1226.0+2954.--} The brightest radio source  
FIRST J122542.2+295616 ($S_{1.4}$ = 13.78 mJy) was
automatically selected by $K$-means as a possible association. 
However, it was later manually discarded as a possible `impostor' since 
its optical counterpart is listed as extended by the SDSS pipeline.  
Analysis of the \fermi\ LAT data shows no high-energy 
photons with energies above 10 GeV.

\noindent
{\bf 1FGL J1515.5+5448.--} In this case, 
the brightest radio source FIRST J151603.0+545629
($S_{1.4}$ = 26.89 mJy) is steep and falls in the neighbourhood of 
a SDSS optical object
flagged as extended. 
FIRST J151444.1+545027 ($S_{1.4}$ = 4.91 mJy)
should also be examined as a potential counterpart.

\noindent
{\bf 1FGL J1527.6+4152.--} At $S_{1.4}$ = 77.77 mJy,
FIRST J152757.5+414708 is the dominant radio source within the
$\gamma$-ray error region. It is also detected by WENSS and GB6, 
but falls
outside the locus of association determined by $K$-means. Further
analysis shows that its spectral index
is rather steep $\alpha_{92} = -1.04$. 
FIRST J152735.2+414839 ($S_{1.4}$ = 2.93 mJy) 
is most likely associated with a pair of interacting galaxies 
at $z=0.149$.

\noindent
{\bf 1FGL J1553.9+4952.--} Three prominent ROSAT sources are
found in or around this region: 1RXS J155357.1+495930
(with a photometric redshift 
$z=0.425$), 1RXS J155437.5+495915 (QSO, $z=0.905$),
and 1RXS J155254.9+495818 (no spectroscopy). However,
none is radio loud.  The brightest radio source inside
FIRST J155234.8+495446 ($S_{1.4}$ = 7.72 mJy) 
reveals a steep radio spectrum $\alpha_{92} = -0.98$ and no 
obvious optical counterpart. 
1FGL J1553.9+4952 has the highest energy $\gamma$-ray photon
(E = 265.4 GeV) detected among the unassociated \fermi\ sources
in the `overlap region'.  

\noindent
{\bf 1FGL J1627.6+3218.--} FIRST J162715.3+321652 ($S_{1.4}$ = 14.68 mJy)
is  steep in radio and remains undetected in the optical by SDSS 
down to a limit of $r > 23.1$. 
We find no \fermi\ LAT
photons with energies above 10 GeV within the error region. Interestingly,
there is a single \fermi\ LAT photon 
with energy E = 10.6 GeV that falls outside the error region 
but may coincide with the radio-quiet ROSAT source 1RXS J162851.9+322655. 
The latter is a rather soft X-ray object with 
a hardness radio of -0.83 that deserves some attention.  

\noindent
{\bf 1FGL J1630.5+3735.--} Inside the error region,
we find FIRST J163056.1+374227 ($S_{1.4}$ = 6.09 mJy) that
shows a steep spectral index $\alpha_{92} = -1.12$. There is an extended
SDSS optical counterpart in the vicinity.  
The  \fermi\ LAT data for this source
is void of individual photons with energies above 10 GeV.

\end{document}